\documentclass[sigconf]{acmart}

\usepackage{booktabs} 
\usepackage{graphicx}
\usepackage[caption=false]{subfig}

\copyrightyear{2018} 
\acmYear{2018} 
\setcopyright{acmcopyright}
\acmConference[PEARC '18]{Practice and Experience in Advanced Research Computing}{July 22--26, 2018}{Pittsburgh, PA, USA}
\acmBooktitle{PEARC '18: Practice and Experience in Advanced Research Computing, July 22--26, 2018, Pittsburgh, PA, USA}
\acmPrice{15.00}
\acmDOI{10.1145/3219104.3229282}
\acmISBN{978-1-4503-6446-1/18/07}

\begin{document}
\title{Discovering Job Preemptions in the Open Science Grid}

\author{Zhe Zhang}
\affiliation{%
  \institution{Holland Computing Center}
  \streetaddress{P.O. Box 1212}
  \city{Lincoln}
  \state{Nebraska}
  \postcode{68588-0150}
}
\email{zhan0915@huskers.unl.edu}

\author{Brian Bockelman}
\affiliation{%
  \institution{Holland Computing Center}
  \streetaddress{P.O. Box 1212}
  \city{Lincoln}
  \state{Nebraska}
  \postcode{68588-0150}
}
\email{bbockelm@cse.unl.edu}

\author{Derek Weitzel}
\affiliation{%
  \institution{Holland Computing Center}
  \streetaddress{P.O. Box 1212}
  \city{Lincoln}
  \state{Nebraska}
  \postcode{68588-0150}
}
\email{dweitzel@cse.unl.edu}

\author{David Swanson}
\affiliation{%
  \institution{Holland Computing Center}
  \streetaddress{P.O. Box 1212}
  \city{Lincoln}
  \state{Nebraska}
  \postcode{68588-0150}
}
\email{dswanson@cse.unl.edu}

\renewcommand{\shortauthors}{Z. Zhang et al.}

\begin{abstract}
The Open Science Grid(OSG)\cite{osg} is a world-wide computing system which facilitates distributed computing for scientific research. It can distribute a computationally intensive job to geo-distributed clusters and process job's tasks in parallel. For compute clusters on the OSG, physical resources may be shared between OSG and cluster's local user-submitted jobs, with local jobs preempting OSG-based ones. As a result, job preemptions occur frequently in OSG, sometimes significantly delaying job completion time.

We have collected job data from OSG over a period of more than 80 days. We present an analysis of the data, characterizing the preemption patterns and different types of jobs. Based on observations, we have grouped OSG jobs into 5 categories and analyze the runtime statistics for each category. we further choose different statistical distributions to estimate probability density function of job runtime for different classes.
\end{abstract}

%
%

\ccsdesc[500]{Computer systems organization~Reliability}
\ccsdesc[500]{Computer systems organization~Availability}

\keywords{OSG, Pilot job, Job failure, Preemption, Failure pattern, Temporal locality, Spatial locality, Job runtime, Distribution, Probability Density Function, Estimation}

\maketitle

\section{Introduction}
\label{sec:introduction}


The OSG is a national cyber-infrastructure consisting of computational resources distributed among tens of independently-run grid sites.  The users of the OSG are organized into large ``virtual organizations" (VOs), each of which utilizes a specific middleware stack to achieve its goals.  For smaller groups or individual scientists, the OSG provides a hosted job submission system.  To coordinate these user jobs with computational resources, OSG uses HTCondor\cite{condor-practice} to perform scheduling and job executions for millions of jobs on the grid resources. A higher-level resource provisioning  system,  GlideinWMS\cite{glideinWMS}, allocates resources to HTCondor using the pilot paradigm. GlideinWMS submits batch jobs to the individual grid-enabled clusters; when scheduled, these \textit{pilot jobs} launch a HTCondor worker node process that connects to the VO's central pool.  The resulting pool overlays the disparate resources from across the grid, presenting a relatively homogeneous HTCondor batch system to the individual scientist.  The individual submits the scientific \textit{payload jobs} into the central VO pool.  The pilot system is an effective way to manage grid resources, as users aren't exposed to the vagaries of different cluster queuing policies and VOs have fine-grained control over the resources each individual receives.

A pilot job running inside a compute cluster is fairly analogous to a virtual machine running on Amazon's EC2; for example, it must detect and manage a physical resource for a dedicated payload job.  Similarly, like a virtual machine, one must be prepared for the pilot to fail without warning - network cuts or hardware failures being a possible source.  However, one failure mode common in the OSG for opportunistic jobs - but relatively uncommon in cloud providers - is batch job preemption. If, for example, a university computing center configures its grid resources to accept jobs both from OSG (pilot jobs) and local researchers, the university might assign higher priority to the local users and enable preemption on the grid jobs. As a result, OSG pilot-jobs are likely to be preempted if there's a spike of local usage.

Unlike Amazon EC2's ``spot pricing,'' there's little indication to the user when a preemption occurred.  If any feedback is provided, it is to the provisioning layer - glideinWMS - not the user scheduling layer.  Given preemptions are often immediate and without warning, they are often indistinguishable from a network connectivity drop.

At low rates, preemption is relatively harmless across the thousands or millions of jobs a user may run.  At higher rates, preemptions may carry severe impacts for the payloads running inside the pilot.  For example, once a disconnect or preemption is detected, the OSG payload jobs have to be rescheduled and started over (checkpointing is relatively rare).  This can significantly delay job completion time; if the same payload is preempted many times, it can even prevent forward progress in a workflow.

Job failures have been actively researched for last decades. The researchers mainly conducted experiments from two aspects, either from user-job perspective or system-hardware perspective. For example, researchers collected failure event logs from systems\cite{Kola:2005:FLD:2138773.2138831} and predict failures based on different component such as memory failures, IO failures and network failures. Other interesting researches\cite{HUEDO2006727} collected user-submitted job logs and estimates the failures from jobs' return code and other status information. Although previous research provides rich set of methodology to discover failures in grids, the pilot-job based infrastructure makes OSG job failures distinct. Unlike a ``white-box'' experiment of collecting system logs, OSG does not have centralized logging scheme to collect hardware failure from all grids. Unlike ``black-box'' experiment of analyzing user-job logs, one does not have the ability to access full user logs from the disparate submit points.  Further complicating the picture, preemption is typically controlled by site scheduling policy: this is often inscrutable from the outside, changes over time, and often impossible to analytically model across the dozens of sites.

To build a picture of pilot-job failures, we will continuously poll the HTCondor pool composed of the pilot jobs and build an in-memory model of the system over time. This helps us to conduct our experiment as a ``grey-box'' where we collect virtual system logs to analyze failures. 

In addition to predicting job failures in distributed grids, estimating job runtime is another extensive part of research for last decades. Researchers have found that correctly estimating job runtime can help scheduler to make smart decision and reduce the average job completion time\cite{Tsafrir:2007:BUS:1263127.1263243}. Some research\cite{cmsjobruntime} has also estimated job runtime for OSG pilot jobs. However, to the best of our knowledge, none of the prior work have considered preemptions across a set of sites.  The uncertainty of occurrence makes preempted jobs' runtime unpredictable, causing prior models to not fit our case. On the other hand, preemptions themselves have their own patterns. For example, a university compute resources that uses desktop cycle scheduling should see a spike of preemptions at the beginning of the workday.  We work to characterizing the preemptions' behaviors and design a dedicated model to better predict payload runtime.

In this paper, we investigate into preemptions in OSG from the jobs collected over an 80-day period. We characterize common patterns of preemptions across OSG clusters while we also enumerate a few unique signatures. We also try several statistical distributions to estimate probability density function of job runtime on OSG.  This work has the following contributions:

\begin{enumerate} 
\item We design a model to interpret OSG jobs' snapshots to job life cycles and demonstrate interesting preemption behaviors.
\item We characterize runtime for different job classes and use statistical distributions to estimate probability density function of job runtime.
\end{enumerate}


\section{Background}
\label{sec:background}

\subsection{Pilot Systems}

Pilot-based systems are widely adopted by grid infrastructures used for scientific research\cite{PanDA}\cite{DIRAC}\cite{AliEn}. A reason for the success is that they provide a straightforward approach to manage the job-to-site matchmaking model (``In which site should I queue my payload job?") historically used in grid computing that required impossibly-accurate modeling of site queues and reliability. The pilot-based systems improve the prior model in two ways: by using \textit{late-binding}, payload jobs are not assigned to sites until resources are available; second, by decoupling the workload specification from the job execution. Late-binding solves issues associated with committing a payload to a certain site before knowing how long until resource are available. Decoupling workload specification and job execution can help payload jobs to select optimal resource cost through matchmaking algorithms.



\begin{figure}[!ht]
\begin{center}
\includegraphics[width=0.4\textwidth]{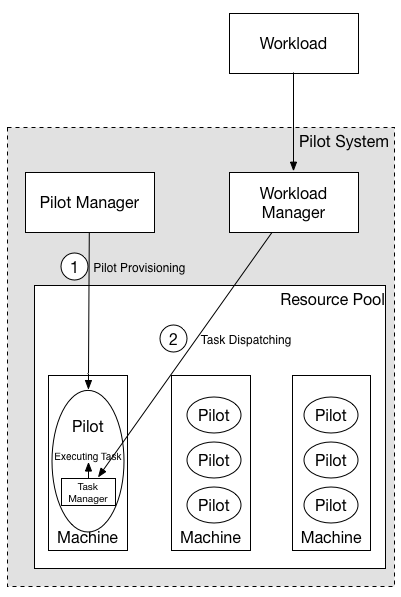}
\end{center}
\caption{High-level abstraction of Pilot Systems}
\label{fig:pilotsystems}
\end{figure}

Figure \ref{fig:pilotsystems} shows a common architecture of a pilot system. To execute a task, the Pilot Manager will first query the resource pool and the workload manager to compare available resources and job demand. As necessary, it may allocate resources by creating a pilot job. The Pilot Manager can also submit these pilots to a remote resource as a job task (a pilot job itself is a job task). When started, the pilot executes daemons on batch worker nodes and is responsible to manage and advertise the resources available for application-related tasks (the payload jobs) to run on the resources. After the resources have been successfully obtained and joined to the resource pool, the Workload Manager will dispatch the payload job to the pilot where the pilot's task manager executes the payload job.



\subsection{GlideinWMS and HTCondor}


For the OSG pilot system studied, the pilot manager is GlideinWMS\cite{glideinWMS} and the workload manager is HTCondor.  In the OSG, this combination is used to connect over 120 individual computing resources across North America and delivers more than 70 million CPU hours a month for scientific research\cite{sharedecosystem}. HTCondor is a well-known workload manager (implying a large user base with good documentation), originally arising from desktop scavenging (implying an emphasis on reliability in an unreliable resource environment).  GlideinWMS provides a mechanism for building a pilot system based on HTCondor -- in addition to using HTCondor throughout its internals.

One advantage of this setup is that the standard HTCondor daemons handle the scheduling of user jobs to available resources.  For this paper, three pertinent roles inside HTCondor are:
\begin{itemize}
\item The 
\textbf{schedd} daemon that manages a job queue where users can submit jobs.
\item The
\textbf{startd} daemon manages a computing resource and executing any corresponding user jobs.  Periodically, it advertises its status to the central manager.
\item On the \textit{central manager}, the
\textbf{collector} daemon collects the advertised state from all daemons in the pool.  The \textbf{negotiator} queries the collector for resources and schedds for jobs, and performs matchmaking.
\end{itemize}

In addition to traditional resource attributes (available CPUs, memory), on the OSG the startd will advertise local information about the grid site name such as its pilot job ID or site name.

When a payload job is matched to a resource (pilot job) by the central manager, the schedd will receive a connection string and a capability string representing a \textit{claim} for running on the resource.  The schedd will contact the startd directly and ``activate" the claim to start job execution.  The startd will advertise its updated state to the central manager and keep a TCP connection open between the pilot and schedd to act as a job heartbeat.  Thus, one can build an approximation of the system's state by solely querying the central manager.

The glideinWMS pilot has three phases of its life:

\begin{itemize}
\item \textbf{Normal running}: During this phase, the pilot behaves like a standard HTCondor worker node \textit{except} that it will shut down if it has been idle (no running payload jobs) for more than 20 minutes.
\item \textbf{Retirement} During the phase, the pilot will accept no new payload jobs and will shut down immediately once there are no running payloads.
\item \textbf{Shutdown} At this point, the pilot will shut down immediately, preempting the payload (causing the remote schedd to reschedule it).
\end{itemize}

The transition from Normal running to Retirement to Shutdown is time-based and determined at pilot job startup.

Some site batch systems configure the pilot to be re-queued after preemption, resulting in multiple instantiations of the same submitted pilot job.  Thus, in the next section, it will be important to distinguish between the unique pilot job in a site batch system and the one-or-more instantiations of that job.

\section{Job Overview on OSG}
\label{sec:joboverview}


In the OSG, the collector is a central resource manager that is responsible to negotiate user jobs with available resources. Each pilot job run by the OSG reports periodically (approximately once every 5 minutes) to the collector. The report includes pilot job information such as the states, activities, run time and planned stop time; this information can be queried by any HTCondor client.



\subsection{Data Collection}



We sample the status of all pilot jobs in the OSG central manager once every minute; for each known pilot, the collector returns a \textit{ClassAd}\cite{matchmaking}, a set of key-expression pairs describing the pilot's state (similar to a JSON map).  For each pilot, we collect the following attributes from the ad:

\begin{itemize}
\item
\textbf{Name}: Pilot name.  Randomly generated when the pilot starts.
\item
\textbf{State}: Current HTCondor startd state; typically ``claimed,'' ``unclaimed,'' or ``retiring.''.
\item
\textbf{Activity}: Current activity of the startd; can be thought of a sub-state. 
\item 
\textbf{MyCurrentTime}: Current Unix timestamp when the collector sent the ad back to the client.
\item
\textbf{TotalJobRunTime}: time, in seconds, since the currently running payload job started.
\item
\textbf{DaemonStartTime}: Unix timestamp when the pilot job was started by the site batch system.
\item
\textbf{GLIDEIN\_ToRetire}: Unix timestamp when the pilot job will start retirement mode (will accept no new payloads).
\item
\textbf{GLIDEIN\_ToDie}: Unix timestamp when pilot will shut down, regardless of whether a payload is running.
\item
\textbf{GLIDEIN\_Site}: Site name where the pilot is running.
\item
\textbf{GLIDEIN\_Entry\_Name}: a unique name (the ``entry point'') describing the combination of the pilot configuration and the cluster where the pilot was submitted.  For example, two pilots at the same site but different batch system queues would have distinct values for GLIDEIN\_Entry\_Name.
\item
\textbf{GLIDEIN\_ResourceName}: The human-readable name of the cluster where the job was submitted (there are multiple resources per site).
\item
\textbf{GLIDEIN\_SITEWMS\_JobId}: The unique pilot job ID assigned when it is submitted to the site.
\end{itemize}

Note the GLIDEIN\_SiteWMS\_JobID is generated by the site batch system when the pilot is submitted and is invariant even if the site preempts and restarts the job; however, the Name and DaemonStartTime changes each time the pilot is started.

\subsection{Data Preprocessing}

Starting with the collected snapshots, we aim to produce a timeline of each pilot job.  Each pilot job is identified by a unique tuple, (GLIDEIN\_SITEWMS\_JobId, GLIDEIN\_Entry\_Name), which persists across multiple pilot preemptions of the same pilot; the pool snapshots are grouped by this unique identifier.  Recorded ClassAds associated with the same pilot are sorted in chronological order.  We then compare each snapshot with its previous snapshot, building a timeline of state transitions.  For example, if a pilot job disappears from one snapshot to the next, we consider its execution to have potentially terminated.

Each time we observe a potential termination, we'd like to to classify the termination reason based on the prior pilot snapshot and any potential future ones (e.g., pilot was restarted or recovered from a network outage).

\begin{figure}[!ht]
\begin{center}
\includegraphics[height=1.5in, width=3.5in]{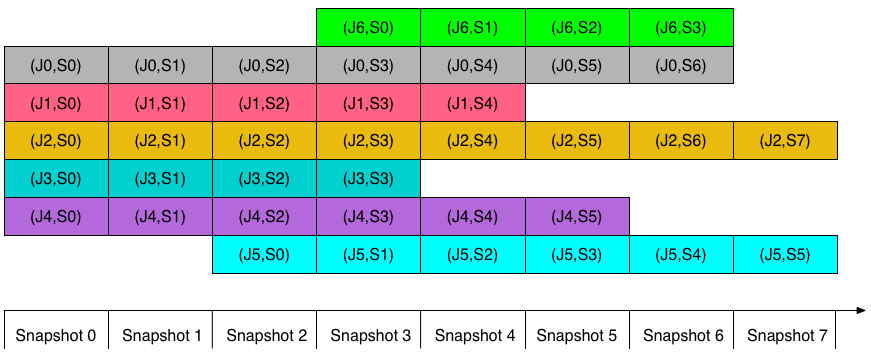}
\end{center}
\caption{Determining Pilot Timelines based on Collector Snapshots}
\label{fig:jobsnapshot}
\end{figure}

Figure \ref{fig:jobsnapshot} shows pilot jobs along time series. There are 8 snapshots. At the first snapshot, Job0, Job1, Job2, Job3 and Job4 are in the OSG central manager and considered running. At the Snapshot2, Job5 starts running and at the Snapshot3, Job6 starts running. If we group jobs in a single snapshot (vertically), we can get the number of running pilots. On the other hand, if we group snapshots for a single job(horizontally), we can visualize a timeline of the pilot. 

Based on the reconstructed pilot timelines, we can infer additional attributes about the pilot. For example, we can estimate the  runtime of a single instantiation of a pilot by subtracting the pilot's MyCurrentTime in the last-seen snapshot by DaemonStartTime.  Each instance can be detected by the Name attribute, which changes at instance start.  The estimated aggregate runtime for a single pilot is then the sum of all estimated pilot instance runtimes.  

\subsection{Job Labeling}

Based on observed and understood pilot job behaviors in OSG, we attempt to classify each potential termination event into one of five categories:

\begin{itemize}
\item \textbf{Retire} - the pilot was in its Retirement phase and the last payload exited.
\item \textbf{Kill} - the pilot hit its Shutdown phase, causing it to immediately preempt any payloads and terminate.
\item \textbf{IdleShutDown} - the pilot is in the Normal phase but shuts itself down due to an idle timeout (no payload jobs were received within a configured time window). 
\item \textbf{Preemption}: the pilot was killed by the site batch system.  Preempted pilots may-or-may-not be restarted in the future by the site batch system, depending on the site configuration, preemption reason, and pilot age.
\item \textbf{NetworkIssue}: the potential termination event was actually a network disconnect; in this case, the same pilot instance will appear in a snapshot in the future.
\end{itemize}



\begin{figure}[!ht]
\begin{center}
\includegraphics[height=2in, width=3in]{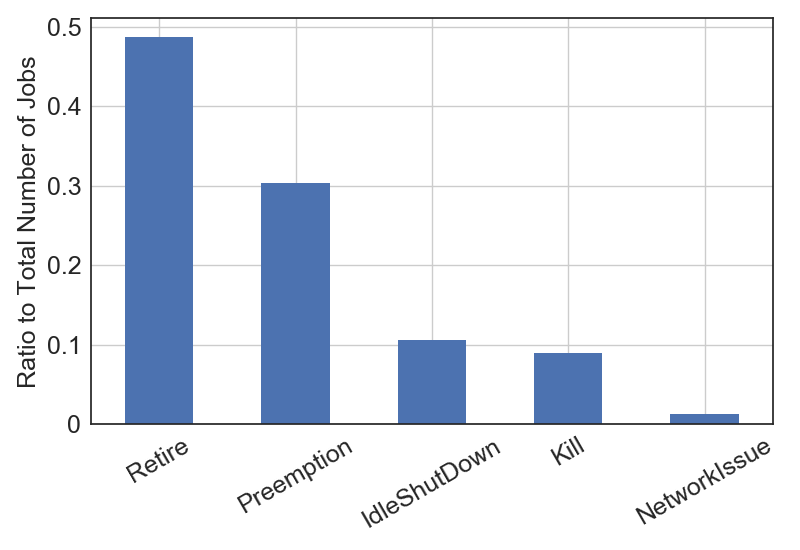}
\end{center}
\caption{Job Distribution on OSG}
\label{fig:classjobdist}
\end{figure}

Note the same pilot's unique identifier can re-appear multiple times in the case of Preemptions (multiple pilot instances) and NetworkIssues (same pilot instance).  In our dataset, we recorded 1,962,608 distinct pilot jobs and 2,314,666 pilot job instances. 
Figure \ref{fig:classjobdist} shows the distribution of different types of instances. As seen in the figure preemptions occur frequently and contribute to 30.37\% of total job instances.

48.64\% are Retire jobs. These jobs indicate that OSG resources maintain a healthy state. 10.63\% are IdleShutDown jobs and these jobs indicate resources are recycled due to decreasing workloads. Only 1.34\% jobs encounters network issues. These jobs temporarily disappear during some snapshot but reappear later. NetworkIssues are different from Preemptions because they keep the same DaemonStartTime when they reappear in snapshots.

\subsection{Pilot Classification over Time}

\begin{figure}[!ht]
\begin{center}
\includegraphics[height=4in, width=3.5in]{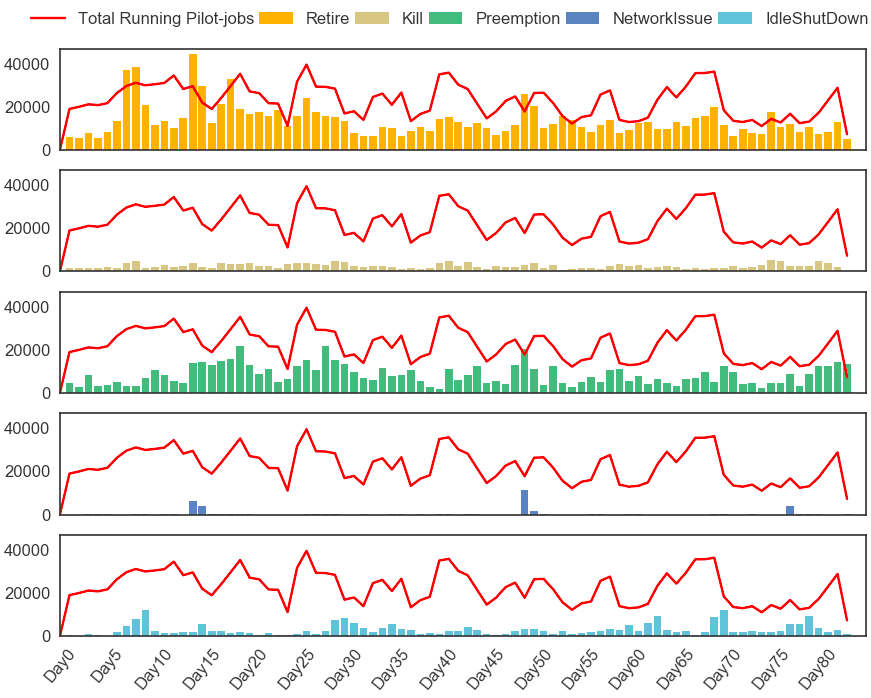}
\end{center}
\caption{Different Pilot Potential Termination Events over Observation Period}
\label{fig:timeseries}
\end{figure}

Figure \ref{fig:timeseries} shows the occurrences of different jobs over our observed time period. As seen in the figure, Preemptions and NetworkIssues occurred in a burst pattern. The overall jobs sharply decrease when Preemptions or NetworkIssues happen. In addition, at some time points Preemptions and NetworkIssues happen at the same time.


\subsection{Pilot Distribution across Different Clusters}

\begin{figure}[!ht]
\begin{center}
\includegraphics[height=2.7in, width=3.5in]{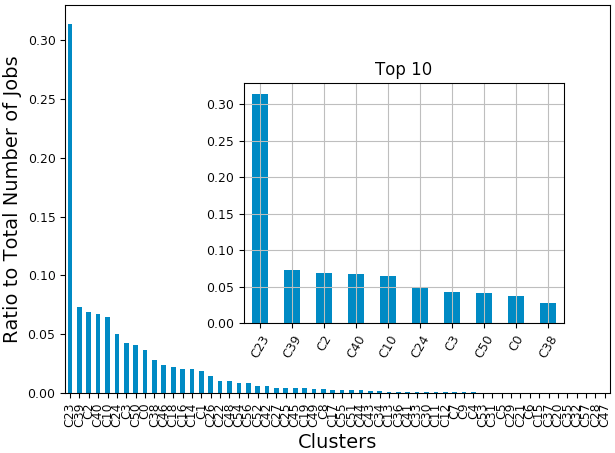}
\end{center}
\caption{Pilot-job Distribution across Different Clusters}
\label{fig:resourcejobdist}
\end{figure}

Figure \ref{fig:resourcejobdist} shows the job distribution on different clusters. As seen in the figure, there is a large variance across different clusters. One of the reasons is due to the variant of configurations on different clusters. For example, some clusters allow preemptions but some clusters do not; different clusters might configure their GLIDEIN\_ToRetire, GLIDEIN\_ToDie to different values and thus jobs are terminated at different duration.

Not all cluster batch systems are configured to actively preemption jobs; we will examine more closely the top sources of preemption.  However, we expect all clusters to have some minimal rate of preemption as other causes of pilot job failure (such as hardware failure or a payload consuming too much RAM) will be indistinguishable from preemption and counted as preemption in our model.

\subsection{Job Duration}

\begin{figure}[!ht]
\begin{center}
\includegraphics[height=2in, width=3in]{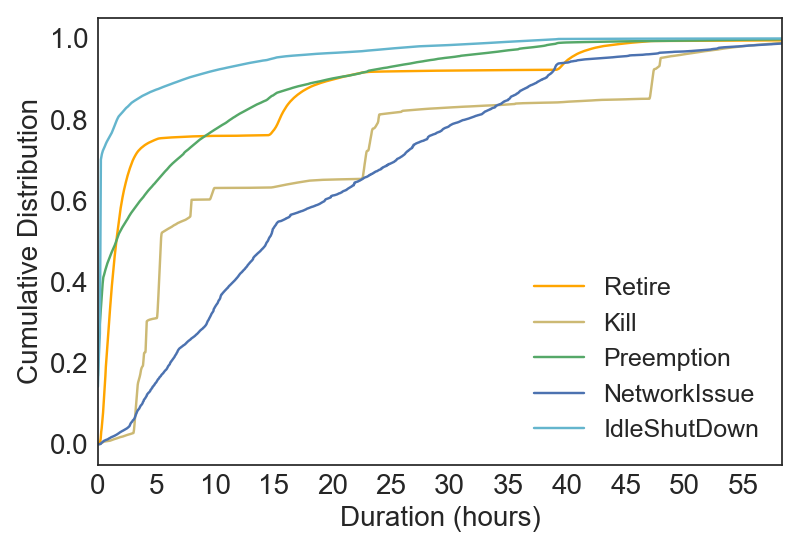}
\end{center}
\caption{Job Duration Distribution of Different Jobs}
\label{fig:durationdistribution}
\end{figure}

Figure \ref{fig:durationdistribution} shows the cumulative distribution function(CDF) of different classes. Preemptions can significantly decrease the pilot job runtime. As seen in Figure \ref{fig:durationdistribution}, 80\% of IdleShutDown pilots run for less than 1 hour. Around 40\% of Preemptions also happen within 1 hour. Retire pilots appear to have several steps in the CDF. We believe those steep increases in CDF are the result of pre-configured GLIDEIN\_ToRetire for pilot jobs. Kill pilots also have similar step-like pattern in CDF. However, comparing to Retire jobs, Kill pilots tend to have even sharper increasing edges. The reason is because the pre-configured GLIDEIN\_ToDie is more strict for pilot jobs than GLIDEIN\_ToRetire. Once a pilot job reaches GLIDEIN\_ToRetire, it can continue working on its existing payload job until work is finished or pilot job reaches pre-configured GLIDEIN\_ToDie. If a pilot job goes beyond GLIDEIN\_ToDie, it will be intermediately killed. This strict time threshold can result in the sharply increasing steps in CDF. NetworkIssues, as illustrated in the figure, have varying pilot job lifetimes. As we pointed out before, NetworkIssues do not terminate pilot jobs and those payload jobs may continue to run to completion.



\section{Observations for Preemptions}
\label{sec:preemption}

\subsection{Preempted jobs are likely to be preempted again}

\begin{figure}[!ht]
\begin{center}
\includegraphics[height=2.5in, width=3.5in]{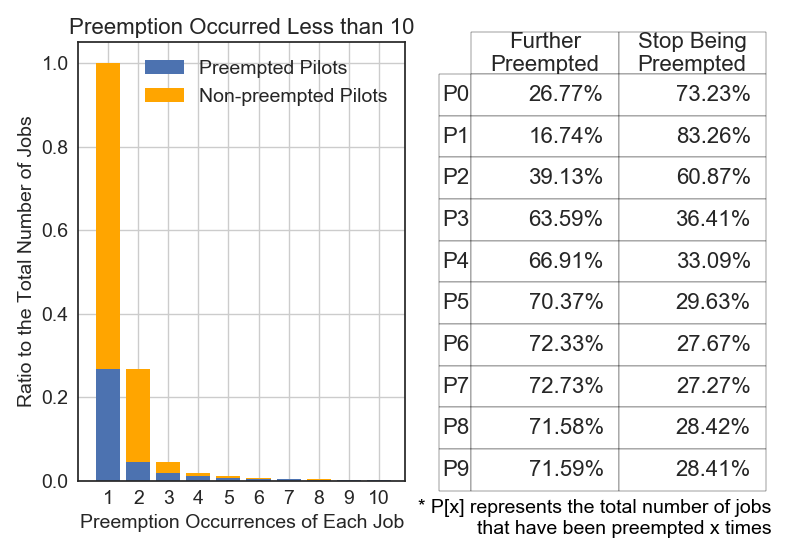}
\end{center}
\caption{Distribution of Preemption Times}
\label{fig:preemptiondist}
\end{figure}

As we observed over 80 days, pilot preemptions are not rare on the OSG. 26.77\% of pilot jobs have encountered at least one preemption. In addition, Preemptions can occur multiple times on the same pilot. In one extreme case, we observed a pilot had been preempted 123 times. Figure \ref{fig:preemptiondist} shows the continuous behavior of preemptions. The left figure in Figure \ref{fig:preemptiondist} shows the ratios of continuously preempted jobs to the total number of jobs in the OSG. As shown in the figure, there are 26.77\% of the pilots have been preempted at least once, 4.48\% of the pilots have been preempted at least twice and so on. As the same jobs continuously get preempted, they become more likely to be preempted again. The right table in Figure \ref{fig:preemptiondist} shows the percentages of continuous preemptions that occur within the preceding preempted jobs. As seen in the table, once the pilots get preempted for the first time, 16.74\% of them get preempted for the second time. Among the pilots that have been preempted twice, 39.13\% gets preempted for the third time. Among those pilots, 63.59\% gets preempted for the fourth time and so on. 

\subsection{Preemptions are correlated with specific grid resources}

Each cluster has its local preemption configuration and local usage patterns. So far, we have measured the OSG as a black-box and manually created a classification based on our experience. Within the 56 clusters we observed, we group them into three categories:
\begin{itemize}
\item Clusters that preempt, but re-queue preempted jobs for another execution.
\item Clusters that ``destructively" preempt; preempted jobs are not re-run.
\item Clusters that do not appear to preempt; if preemptions occur, they are either misclassified or lack enough OSG pilots for us to identify the preemptions.
\end{itemize}

\begin{figure}[!ht]
\begin{center}
\includegraphics[height=2.7in, width=3.5in]{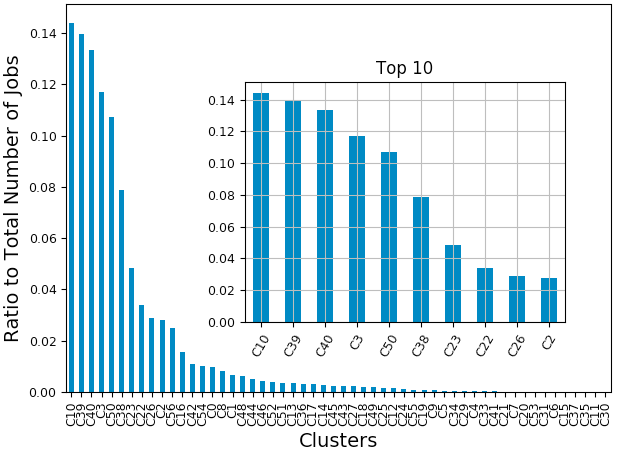}
\end{center}
\caption{Preemptions on Different Grid Resources (Top 10)}
\label{fig:sitepreemptdist}
\end{figure}

Figure \ref{fig:sitepreemptdist} shows preemption rates for all cluster resources where any preemptions were detected. To de-identify the clusters, we replace the human-readable names with an enumeration prefixed with \texttt{C}. Note the preemption distribution is different from the pilot distribution shown in Figure \ref{fig:resourcejobdist}.

\subsection{Preemptions Occur Early in Job Runtime}

\begin{figure}[!ht]
\begin{center}
\includegraphics[height=2in, width=3in]{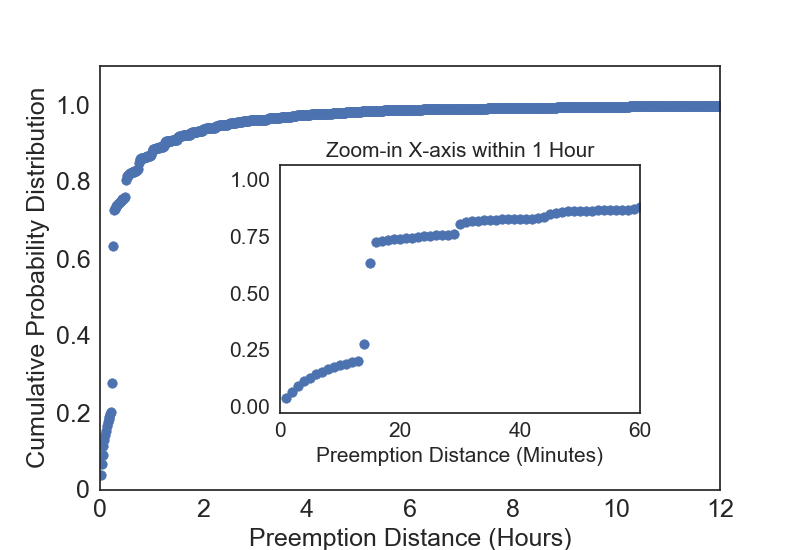}
\end{center}
\caption{Time-to-preemption CDF on C39}
\label{fig:suogcepreemptiondistance}
\end{figure}

Figure \ref{fig:suogcepreemptiondistance} depicts the CDF of preemptions on cluster C39. Here, preemptions occurs in bursts: over 80\% of preemptions happens within 1-hour from the pilot instance start. 

\section{Probability Density Function}
\label{sec:pdf}

In this section, we present probability distribution of different job classes. We try to find common characteristics of a specific job class across different clusters. Our purpose is to estimate the probability density function of job runtime for a cluster. We run the following tests on every cluster but only present results about C10 (carrier most jobs in OSG) due to the space limit of the paper. We also give general summary of other clusters in the end.

To estimate runtime distribution, we use Normal distribution, Uniform distribution, Gamma distribution, Chi-squared distribution, Johnson $S_U$(Unbounded) distribution\cite{jobruntime}, Johnson $S_B$(Bounded) distribution, Inverted weibull distribution and Exponential weibull distribution\cite{exponentialweibull}. Normal, Uniform, Gamma and Chi-squared distributions are well known distributions. Johnson $S_U$ and $S_B$ distributions belongs to Johnson's distribution system which was developed by Johnson in 1949. It is a flexible system of distributions, based on three families of transformations, that translate an observed, non-normal variate to one conforming to the standard normal distribution. The exponential, logistic, and hyperbolic sine transformations are used to generate log-normal (SL), unbounded (SU), and bounded (SB) distributions, respectively. Inverted and Exponential Weibull distributions belongs to weibull distribution which is a popular statistical distribution that is used to estimate important life characteristics of the product such as reliability or probability of failure at a specific time, the mean life and the failure rate. We use these 8 distributions and try to fit runtime distributions of different job classes.

\subsection{PDF Estimation on C10}

\begin{figure}[!ht]
\begin{center}
\includegraphics[height=2.5in, width=3.5in]{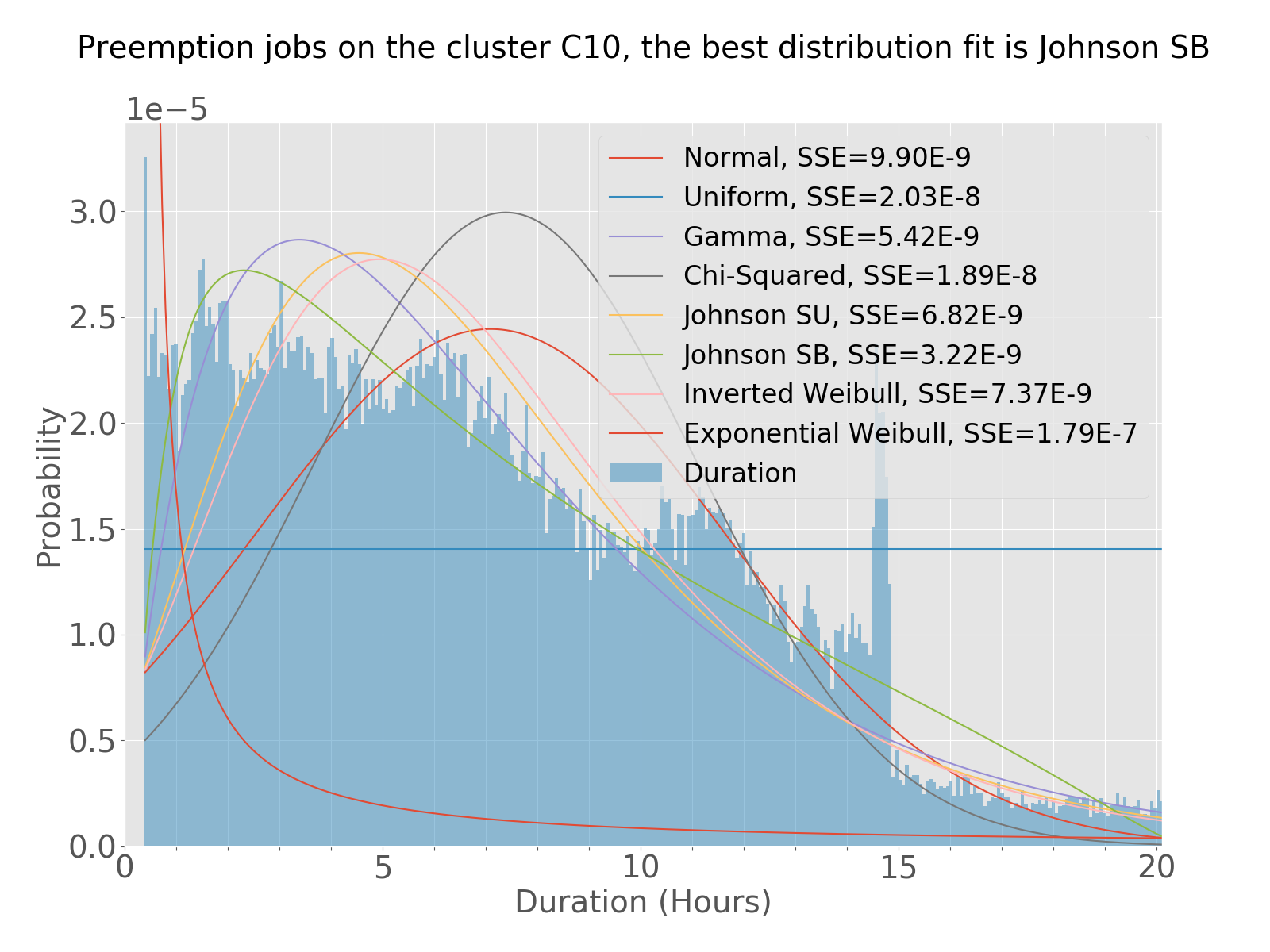}
\end{center}
\caption{PDF Estimation of Preemptions on C10}
\label{fig:preemptionglowpdf}
\end{figure}

Figure \ref{fig:preemptionglowpdf} shows the job runtime distributions of C10. As seen in the figure, Johnson $S_B$ is the best fit distribution. In top 10 list of clusters with most preemptions in OSG. 5 clusters can be described by Johnson $S_B$ and 4 clusters are fitted by Inverted Weibull and 1 cluster is best fitted by Exponential weibull.

\begin{figure}[!ht]
\begin{center}
\includegraphics[height=2in, width=3in]{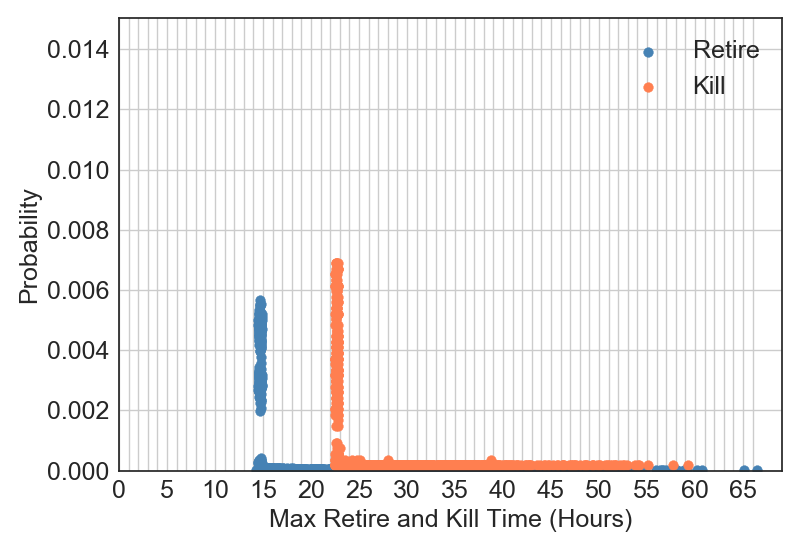}
\end{center}
\caption{Distribution of MaxRetireTime and MaxKillTime on C10}
\label{fig:timemaxglowpdf}
\end{figure}

Figure \ref{fig:timemaxglowpdf} shows distributions of MaxRetireTime and MaxKillTime on C10. As we expected, most of the pre-configured retire and kill time are located at certain time. MaxRetireTimes are located around 15 hours and MaxKillTime are located around 23 hours.

\begin{figure}[!ht]
\begin{center}
\includegraphics[height=2.5in, width=3.5in]{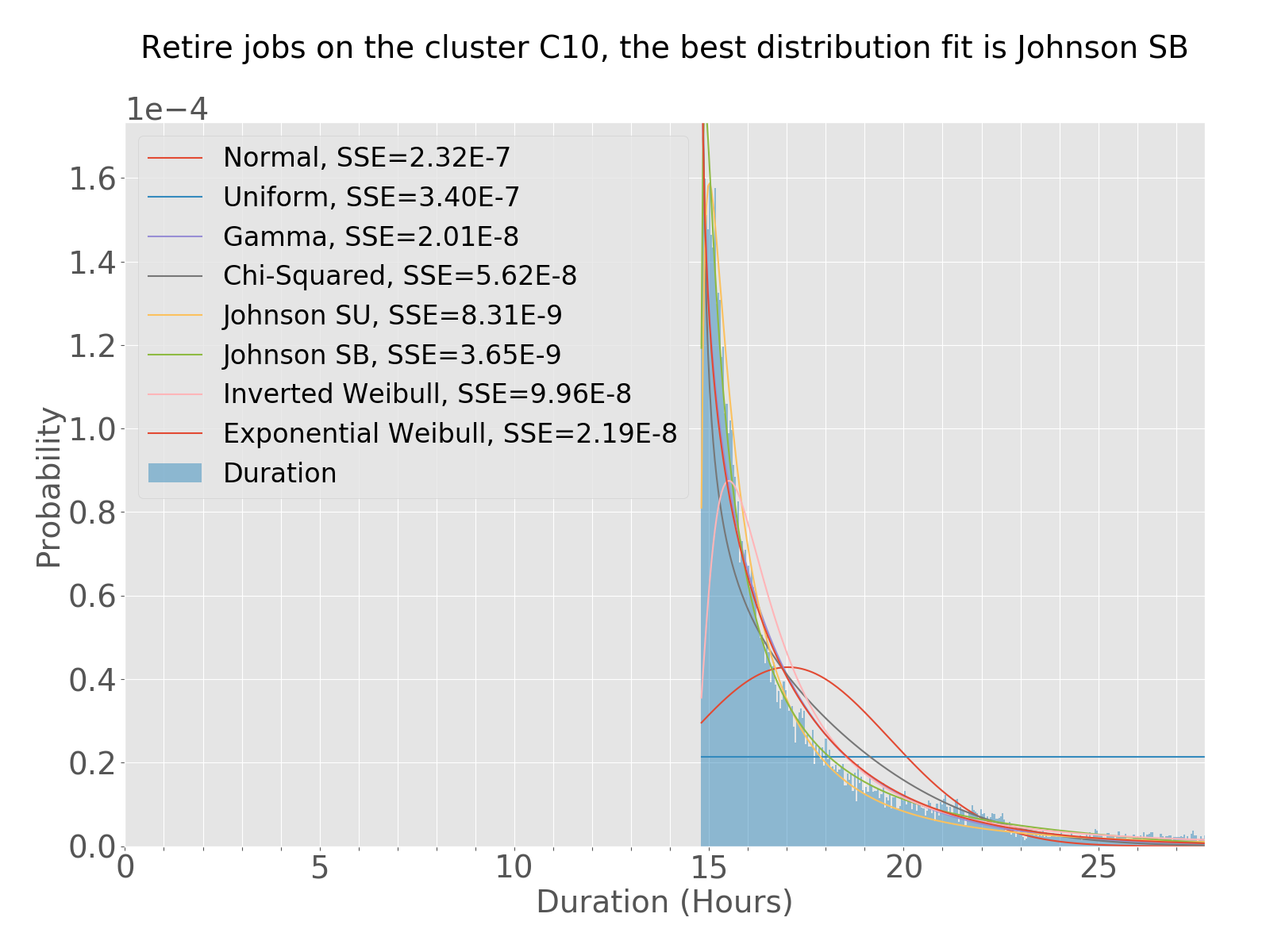}
\end{center}
\caption{PDF Estimation of Retire Jobs on C10}
\label{fig:retireglowpdf}
\end{figure}

\begin{figure}[!ht]
\begin{center}
\includegraphics[height=2.5in, width=3.5in]{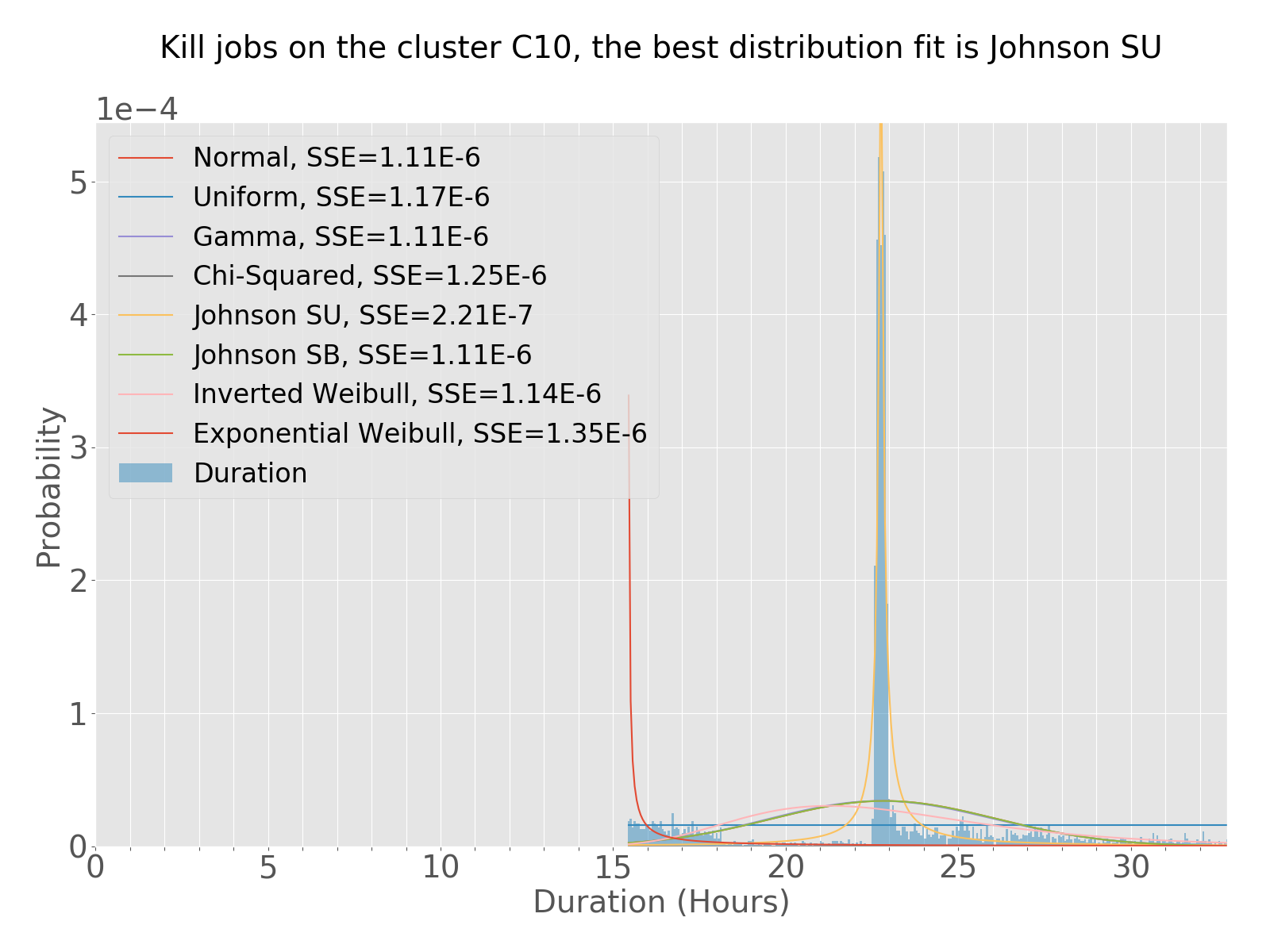}
\end{center}
\caption{PDF Estimation of Kill Jobs on C10}
\label{fig:killglowpdf}
\end{figure}

Figure \ref{fig:retireglowpdf} and Figure \ref{fig:killglowpdf} show runtime distribution of Retire jobs and Kill jobs. As seen in two figures, most of Retire jobs' runtime are around 15 hours and Kill jobs' runtime are around 23 hours. However, comparing to Kill jobs, Retire jobs have longer tails in distribution. This is because Retire jobs have been assigned a ``soft-deadline"(MaxRetireTime) and pilot-jobs can continue to run their existing pillow-jobs until MaxKillTime reached. In this scenario, MaxKillTime is considered as a ``hard-deadline" and all pilot-jobs that are still running on the cluster should be immediately terminated. As a result, Kill jobs' runtime distribution looks sharper than Retire jobs.

We also estimate Retire and Kill jobs on other clusters. All 10 clusters with most Retire jobs and Kill jobs are best fitted by Johnson-family distributions.

We use same method to estimate PDFs of NetworkIssues and IdleShutDowns. We found NetworkIssues randomly distributed on runtime. IdleShutDowns tend to be distributed towards small runtime since most of them are terminated at early stage as we discovered in Section \ref{sec:joboverview}. But due to the limited number of these two types of jobs, we do not show the results in the paper.

\begin{figure}[!ht]
\begin{center}
\includegraphics[height=2.5in, width=3.5in]{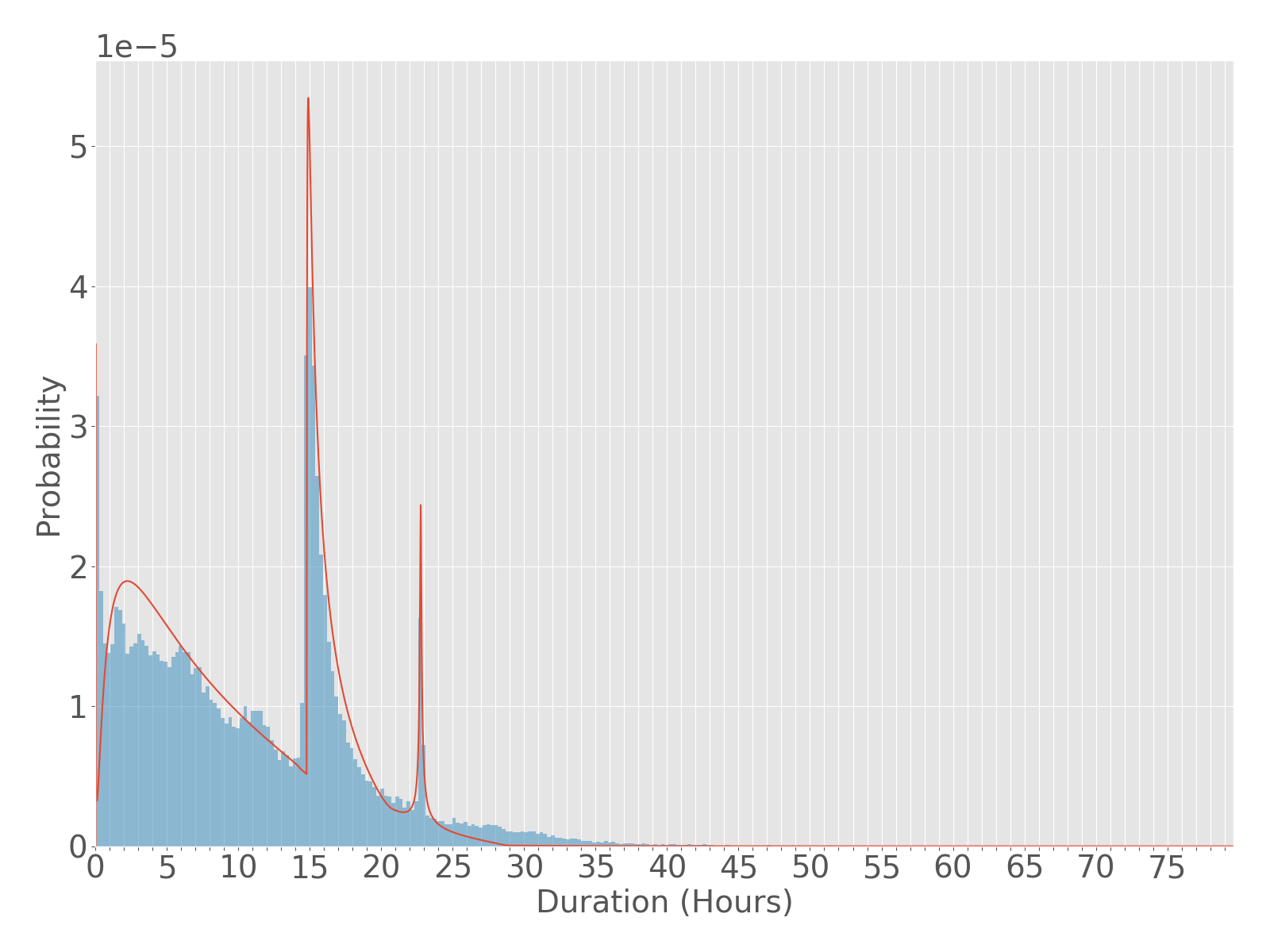}
\end{center}
\caption{Estimation of Overall PDF on C10}
\label{fig:estimationpdfglow}
\end{figure}

With PDFs of different jobs, we further estimate overall PDF of all pilot jobs on C10. Figure \ref{fig:estimationpdfglow} shows the estimated distribution of 5 job classes as well as actual pilot-job distribution.

\section{Conclusions and Future Work}
\label{sec:conclusion}

This paper presents first results from a dataset that allows for analysis of the glideinWMS/HTCondor-based pilot system used on the OSG. In particular, we take a deep look at patterns of preemption for running pilots based on monitoring from the pilots themselves -- we did not rely on ``correct" site behavior for our work.  The approach is important as the plethora of site batch systems and configurations (not necessarily externally queryable) has historically precluded direct analysis of the system's behavior.

We illustrated the temporal and spatial properties of preemptions on cluster sites and also demonstrated the statistical relations between pre-configured system settings(a pilot's to-retire time and to-die time) and actual pilot terminations. All these information together show the ability to estimate the pilot runtime. We believe it will assist us in future scheduling of the payload jobs. In addition to maximizing total payload throughput, we believe this could be used to help guarantee progress on payloads by better matching payload length with remaining pilot runtime.

Additionally, in the papers\cite{Tsafrir:2007:BUS:1263127.1263243}\cite{Sonmez:2009:TEJ:1551609.1551632}, authors use time-series prediction models to job runtime in grids. These models can be generalized to different grids. We plan to try these models in OSG and further improve our prediction performance. 

Finally, we also need to collect more job data from OSG at different time periods such that we can verify if the runtime distributions on a cluster is stationary across different time periods.

\begin{acks}
This work was supported by NSF award PHY-1148698, via subaward from University of Wisconsin-Madison. This research was done using resources provided by the Open Science Grid, which is supported by the National Science Foundation and the U.S. Department of Energy's Office of Science. This work was completed utilizing the Holland Computing Center of the University of Nebraska.
\end{acks}

\bibliographystyle{ACM-Reference-Format}
\bibliography{bib/references}

\end{document}